\documentclass[preprint,12pt]{elsarticle}
\usepackage{amssymb}
\usepackage{float}
\usepackage{subfigure}
\usepackage{amsmath}

\begin{document}

\begin{frontmatter}



\title{CPU-GPU Heterogeneous Code Acceleration of a Finite Volume Computational Fluid Dynamics Solver}

 \author[label1,label3]{Weicheng Xue}
 \affiliation[label1]{organization={Department of Computer Science and Technology, Tsinghua University},
             city={Beijing},
             postcode={100084},
             country={China}}

 \author[label2,label3]{Hongyu Wang}
 \affiliation[label2]{organization={Internet Based Engineering},
             city={Beijing},
             postcode={100083},
             country={China}}             

 \author[label3]{Christopher J. Roy}
 \affiliation[label3]{organization={Department of Aerospace and Ocean Engineering, Virginia Tech},
             city={Blacksburg},
             postcode={24060},
             country={USA}}


\begin{abstract}
This work deals with the CPU-GPU heterogeneous code acceleration of a finite-volume CFD solver utilizing multiple CPUs and GPUs at the same time. First, a high-level description of the CFD solver called SENSEI, the discretization of SENSEI, and the CPU-GPU heterogeneous computing workflow in SENSEI leveraging MPI and OpenACC are given. Then, a performance model for CPU-GPU heterogeneous computing requiring ghost cell exchange is proposed to help estimate the performance of the heterogeneous implementation. The scaling performance of the CPU-GPU heterogeneous computing and its comparison with the pure multi-CPU/GPU performance for a supersonic inlet test case is presented to display the advantages of leveraging the computational power of both the CPU and the GPU. Using CPUs and GPUs as workers together, the performance can be improved further compared to using pure CPUs or GPUs, and the advantages can be fairly estimated by the performance model proposed in this work. Finally, conclusions are drawn to provide 1) suggestions for application users who have an interest to leverage the computational power of the CPU and GPU to accelerate their own scientific computing simulations and 2) feedback for hardware architects who have an interest to design a better CPU-GPU heterogeneous system for heterogeneous computing.
\end{abstract}


\begin{highlights}
\item This work proposed a performance model for CPU-GPU heterogeneous computing to estimate the performance of a finite volume computational fluid dynamics solver. Also, the performance model can provide some guidance for future computer architecture design and domain application performance optimizations.
\item This work applied CPU-GPU heterogeneous computing to accelerate a computational fluid dynamics test case and compared its performance with a pure GPU implementation. With a proper workload ratio assigned to the GPU and CPU workers, the CPU-GPU heterogeneous version outperforms the pure GPU version by up to 23\%, which can be estimated fairly by the heterogeneous computing performance model proposed in this work.
\end{highlights}


\end{frontmatter}


\section{Introduction}

The increasing demand for high-performance computing from computationally intensive applications, including scientific computations and artificial intelligence applications, has greatly driven the development and adoption of heterogeneous computing. As of June 2023 (the latest release), most supercomputers in the Top500 list~\cite{top500} and the Green500 list~\cite{green500} adopt heterogeneous architectures, i.e., different types of processors, accelerators, or specialized hardware. Heterogeneous computing usually combines the advantages of various heterogeneous components, allowing application users to flexibly select appropriate compute units to execute specific computational tasks to satisfy different needs including improving performance, promoting scalability, saving energy, etc. Mittal et al.~\cite{mittal2015survey} surveyed a lot of heterogeneous computing techniques mainly for CPU-GPU heterogeneous computing including the architecture, scheduling, data transfers, programming models, and application domains. They pointed out that CPU-GPU systems are the most important system for heterogeneous computing and matching algorithmic requirements to features of PUs is very important to the performance.

Based on earlier work~\cite{xue2020heterogeneous}, this work is focused on leveraging the total computational power of the CPU and the GPU to achieve high performance for a finite volume Computational Fluid Dynamics (CFD) solver on CPU-GPU heterogeneous systems. CFD usually involves compute-intensive computations and exhibits the feature of single instruction multiple data (SIMD) or single instruction multiple threads (SIMT) since similar calculations are executed for a large number of degrees of freedom (finite cells, finite elements, nodes, etc.). A conventional way of CPU-GPU heterogeneous computing is offloading the compute-intensive portion of CFD computation to the GPU while keeping the rest of CFD work on the CPU~\cite{xue2021improved}, which is one important reason why heterogeneous computing system is developed: assigning appropriate computational tasks to the most suitable components. Different from the conventional way, this work treats both the CPU and the GPU as suitable compute components for the residual calculation in the CFD code, assigning appropriate amounts of workloads to both the CPU and the GPU workers according to their own speeds. The initiative is to prevent the CPU from idling in the conventional mode, making full use of the total computational resources in a single computing system.

Full heterogeneous implementations which utilize modern CPUs and GPUs as equal compute units are relatively rare. Also, there are no appropriate performance models to fairly evaluate the performance of CPU-GPU heterogeneous computing. Alvarez et al.~\cite{alvarez2021hierarchical} presented a hierarchical heterogeneous implementation focusing on three algebra-based kernels on hybrid supercomputers. In their approach, the workload can be distributed among CPUs and GPUs, through the use of CPU computing threads, GPU management threads, CPU management threads and CPU communication threads. CPU computing threads are properly bound to NUMA nodes. Multi-level overlap of communication and computation is shown to be very useful for computing patterns requiring halo exchanges. However, their heterogeneous computing approach has not been proven to be effective for a complete CFD simulation which is usually more complicated. Wang et al.~\cite{wang2018performance} implemented an “MPI+OpenMP+offload” heterogeneous framework for large-scale CFD flow field simulations on a heterogeneous CPU-MIC system, with lots of CPUs and MIC coprocessors being employed. In their framework, each process can offload some share of its work to the corresponding MIC devices and the authors scaled their work to half of the Tianhe-2 supercomputer system. However, the optimizations proposed in their work are tailored to the CPU+MIC architectures (hardware-specific), limiting its direct applicability to other heterogeneous computing systems including the CPU+GPU platform. Although the roofline performance model~\cite{williams2009roofline} can be used as guidance on how to improve the performance of parallel software and hardware by overcoming multiple layers of "performance ceiling", the roofline performance model cannot be used to estimate the advantages of CPU-GPU heterogeneous computing over pure GPU computing.

\section{Finite Volume CFD Solver: SENSEI}

SENSEI is an acronym for Structured Euler Navier-Stokes Explicit Implicit solver. SENSEI was initially developed by Derlaga et al~\cite{derlaga2013sensei}, and extended to a turbulence modeling code base by Jackson et al.~\cite{jackson2019turbulence} and Xue et al.~\cite{xue2020code}. SENSEI adopts multi-block structured grid to ensure mesh quality for general CFD applications. The governing equations for Navier-Stokes equations can be written in a weak form as 
\begin{equation}
\label{governing}
\frac{\partial }{\partial t}\int_\Omega \vec{Q}{\rm d}\Omega +\oint_{\partial \Omega} (\vec{F_{i,n}}-\vec{F_{\nu,n}}){\rm d}A= \int_\Omega \vec{S}{\rm d}\Omega
\end{equation}
where $\vec{Q}$ is the vector of conserved variables, $\vec{F_{i,n}}$ and $\vec{F_{\nu,n}}$ are the inviscid and viscous flux normal components (the dot product of the 2nd order flux tensor and the unit face normal vector), respectively, given as
\begin{equation}
\vec{Q}=
\begin{bmatrix}
\rho\\\rho u\\\rho v\\\rho w\\\rho e_t
\end{bmatrix}, \,
\vec{F_{i,n}}=
\begin{bmatrix}
\rho V_n\\\rho u V_n + n_x p\\\rho v V_n + n_y p\\\rho w V_n + n_z p\\\rho h_t V_n
\end{bmatrix}, \,
\vec{F_{\nu,n}}=
\begin{bmatrix}
0\\n_x \tau_{xx} + n_y \tau_{xy} + n_z \tau_{xz}\\n_x \tau_{yx} + n_y \tau_{yy} + n_z \tau_{yz}\\n_x \tau_{zx} + n_y \tau_{zy} + n_z \tau_{zz}\\
n_x \Theta_{x} + n_y \Theta_{y} + n_z \Theta_{z}
\end{bmatrix}	    
\end{equation}
$\vec{S}$ is the source term from either body forces, chemistry source terms, or the method of manufactured solutions~\cite{oberkampf2010verification}. $\rho$ is the density, $u$, $v$, $w$ are the Cartesian velocity components, $e_t$ is the total energy, $h_t$ is the total enthalpy, $V_n = n_x u+ n_y v + n_z w$ and the $n_i$ terms are the components of the outward-facing unit normal vector. $\tau_{ij}$ are the viscous stress components based on Stokes's hypothesis. $\Theta_i$ represents the heat conduction and work from the viscous stresses.

\section{Discretization}

\subsection{Finite Volume Discretization}

Applying finite-volume discretization, the spatial computational domain $\Omega$ is partitioned into a number of finite control volumes $\Omega_{i}$, so that $\Omega = \bigcup_{i=1}^{N_{\nu}}\Omega_{i}$, where $N_{\nu}$ is the number of finite control volumes and the subscript $i$ is the finite control volume index. The weak form in Eq.~\ref{governing} can be rewritten for each finite control volume as
\begin{equation}
\frac{\partial }{\partial t}\int_{\Omega_{i}} \vec{U}{\rm d}\Omega +\oint_{\partial \Omega_{i}} (\vec{F_{i,n}}-\vec{F_{\nu,n}}){\rm d}s= \int_{\Omega_{i}} \vec{S}{\rm d}\Omega.
\label{governing_cv}
\end{equation}
Denote the discrete solution of the finite volume method as $\Vec{U}_{h}$ which is assumed to be constant in each control volume and approximates the control volume average of the exact solution. The weak form of the general conservation law can be obtained by integrating the conservation law over each of the control volumes and applying the Gauss divergence theorem to the volume integration of the flux divergence term to obtain a surface integral. With the discrete steady-state residual given as $\vec{R}_{h}$, the discrete version of Eq.~\ref{governing_cv} can be given in a semi-discrete form as
\begin{equation}
  |\Omega_{i}|\dfrac{\partial}{\partial t}\vec{U}_{h} + \Vec{R}_{h} = \vec{0},
  \label{spatial_residual}
\end{equation}
where $|\Omega_{i}|$ is the volume of $\Omega_{i}$, $\Vec{U}_{h}$ is the cell averaged solution vector, $\Vec{R}_{h}$ is the spatial residual vector, which is given in Eq.\ref{residual}.
\begin{equation}
  \Vec{R}_{h} = \sum_1^{f} (\vec{F_{i,n}}-\vec{F_{\nu,n}}){\rm \Delta}s - |\Omega_{j}| \vec{S_h}, 
  \label{residual}
\end{equation}
where $f$ is the cell face number ($f = 4$ and $f = 6$ for 2D and 3D, respectively), $\Delta s$ is the face area, and $S_{h}$ is the cell averaged source term vector. The discretized equations in Eq.~(\ref{residual}) can be marched in time using an ODE solver if stability conditions are satisfied.

\subsection{Temporal Discretization}

In this work, second-order explicit Runge-Kutta~\cite{ascher1997implicit, kennedy2016diagonally, jameson1981numerical} 2-step and 4-step temporal schemes are mainly applied. The temporal discretization deals with the numerical approximation of the $\dfrac{\partial}{\partial t}\vec{U}_{h}$ term in Eq.~\ref{spatial_residual}. A conventional form of the Runge-Kutta schemes can be seen in Eq.~\ref{runge_kutta}:
\begin{align}
\begin{split}
  \vec{U}_{h}^{n+1} = & \vec{U}_{h}^{n} - \Delta t \sum_{i=1}^{s} b_i \Vec{R}_{h}^{i}, \\
  \vec{U}_{h}^{i} = & \vec{U}_{h}^{n} - \Delta t \sum_{j=1}^{s} a_{ij} \Vec{R}_{h}^{j},
  \label{runge_kutta}
\end{split}
\end{align}
where $s$ is the number of stages, $b_i$ and $a_{ij}$ are constants, which can be found in Ref.~\cite{ferracina2008strong}. For explicit Runge-Kutta schemes, all $a_{ij} = 0$ when $i \leqslant j$.

\section{Workflow for the GPU-GPU Heterogeneous Computing in SENSEI}

The GPU usually has more lightweight compute cores than the CPU so the compute throughput of the GPU is higher. Besides, the GPU has higher memory bandwidth and lower memory latency to its own memory. These advantages make the GPU more appropriate for compute-intensive computations. However, the CPU can handle versatile tasks including caching, branch prediction, fast single-threaded computation, etc. Their own advantages make each of them supplementary components in modern computing systems. The GPU and the CPU in a system usually have discrete memories so their own data are stored separately, which requires data exchanges between the CPU and the GPU.

\begin{figure}[H]
	\centering
	\includegraphics[width=.8\textwidth]{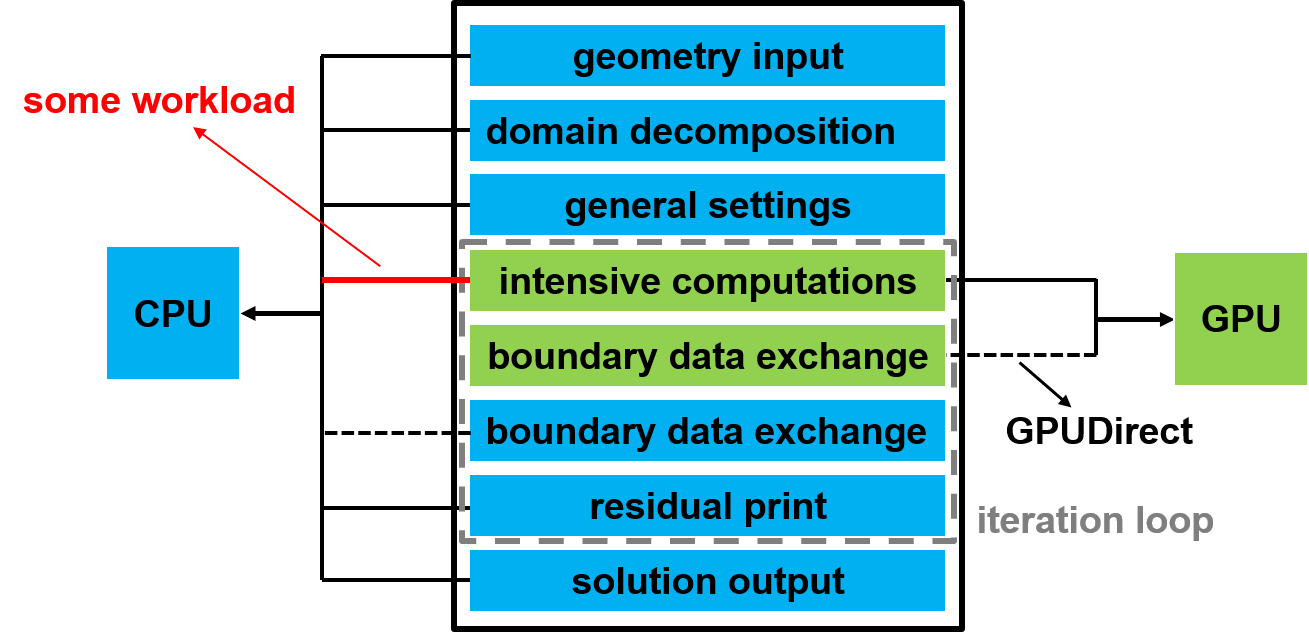}
	\caption{Workflow for the synchronous heterogeneous computing}
	\label{flowchart}
\end{figure}

The workflow for the synchronous heterogeneous computing in SENSEI is depicted in Fig.~\ref{flowchart}. The CPU mainly deals with the geometry input, domain decomposition, general CFD settings, residual print, solution output and a small fraction of computations. Most computations are done on GPU workers with a small amount of workload assigned to CPU workers according to their relative speed. Boundary data exchange can happen between CPU workers or between GPU workers, depending on whether GPUDirect is applied or not. It should be noted that there is a global synchronization among all CPU workers and GPU workers after the boundary enforcement for all boundaries for each iteration substep using RK time marching schemes.

\section{Performance Optimizations using OpenACC Directives}
\label{optimizations}

Some general guidance exists in terms of improving the performance of a program in a heterogeneous system. First, sufficient parallelism should be explored to increase the parallel speedup, that is, the speedup for the parallel portion should compensate for the overhead of data transfers and the parallel setup. Second, the memory or network bandwidth should be improved, which is mainly determined by the data access patterns, data size and data exchange frequency. Some important performance optimizations using OpenACC directives are applied to improve the performance of the CPU-GPU heterogeneous computing in this work. 

\paragraph{Coalesced memory access}
This performance optimization is to improve the memory throughput of the boundary condition data on each GPU worker. Although the GPU worker has higher compute capacity compared to the CPU worker, the CPU worker easily encounters the low memory bandwidth issue. The optimization is targeted at solving this issue by converting the non-contiguous data into a temporary contiguous array in parallel using \texttt{loop for} directives and then updating this temporary array between hosts and devices using \texttt{update} directives.

\paragraph{Ghost cell extrapolation on the GPU}
Ghost cell extrapolation was initially executed on the CPU host, which requires ghost cell exchange between the host and the GPU device. One principle of getting the GPU to work efficiently is keeping the data on the GPU as long as possible and avoiding frequent data exchange between the host and the device. In this optimization, all the ghost cell extrapolation for the GPU is moved to the GPU, instead of execution on the CPU.

\paragraph{Performance Optimization on flux limiter calculation}
SENSEI applies MUSCL extrapolation to reconstruct fluxes, which is given in Eq.~\ref{flux}.

\begin{align}
\label{flux}
\vec{Q}_{i+1/2}^L = & \vec{Q}_i + \frac{\epsilon}{4} [(1 - \kappa) \Psi_{i-1/2}^{+} (\vec{Q}_{i} - \vec{Q}_{i-1}) + (1 + \kappa) \Psi_{i+1/2}^{-} (\vec{Q}_{i+1} - \vec{Q}_{i})] \\
\vec{Q}_{i+1/2}^R = & \vec{Q}_{i+1} - \frac{\epsilon}{4} [(1 + \kappa) \Psi_{i+1/2}^{+} (\vec{Q}_{i+1} - \vec{Q}_{i}) + (1 - \kappa) \Psi_{i+3/2}^{-} (\vec{Q}_{i+2} - \vec{Q}_{i+1})]
\end{align}
where $\epsilon$ and $\kappa$ are MUSCL extrapolation parameters, $\Psi$ are limiter function values. $L$ and $R$ denote the left and right states, respectively.

In the unoptimized version, the limiter variable $\Psi$ is set to be private and local to every CUDA thread, i.e., different CUDA threads have their own copies of limiters on the left and right faces to avoid thread contention. However, this requires the limiter to be computed 2 times redundantly compared to the limiter calculation on the CPU. To fix this issue, limiters are stored in a global array and the calculation of limiters is separated from the flux reconstruction.

\section{A Performance Model for CPU-GPU Heterogeneous Computing}

In this work, a performance model for CPU-GPU heterogeneous computing is proposed to estimate and predict the performance utilizing both the CPU and GPU as workers.

\begin{figure}[H]
	\centering
	\includegraphics[width=.7\textwidth]{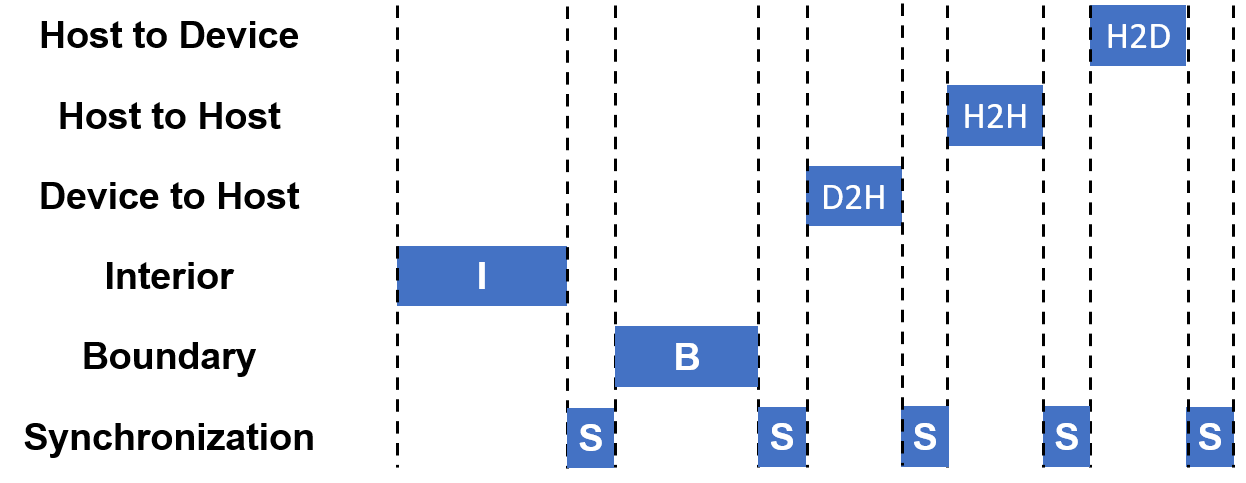}
	\caption{An abstraction of CPU-GPU heterogeneous computing in SENSEI}
	\label{workflow}
\end{figure}

For CFD computations using SENSEI or similar CFD code bases, important computational procedures can be abstracted into high-level computation and communication patterns and chronologically organized into a workflow chart, shown in Fig.~\ref{workflow}. One single iteration of computation is divided into one interior domain residual calculation, followed by the boundary condition enforcement and some data transfers including device-to-host data transfer, host-to-host data transfer and host-to-device data transfer. Therefore, a single iteration of the residual calculation can be broken into 5 main components, each of which is followed by one synchronization to guarantee the former task is finished. It should be noted that the overhead for these different synchronizations may not be the same for a real application, but their effects can be aggregated into a total variable, such as a sequential fraction.

For a cubic domain with $N_l \times N_w \times N_d$ interior cells, the sequential execution time to finish one iteration step on a single CPU is given in Eq.~\ref{sequential}.

\begin{equation}
\label{sequential}
time_{seq} = N_l N_w N_d t_I + (2 N_l N_w + 2 N_l N_d + 2 N_w N_d) t_B + t_{DH} + t_{HH} + t_{HD} + 5 t_S
\end{equation}
where $t_I$ is the execution time to finish one iteration for one interior domain cell on the CPU, $t_B$ is the execution time to finish one iteration for one boundary domain cell, $t_S$ is the synchronization time, $t_{DH}$, $t_{HH}$ and $t_{HD}$ represent the time for device-to-host data transfer, host-to-host data transfer and host-to-device data transfer, respectively.

For a 2D CFD test case using 1D domain decomposition, $N_d = 1$. For CPU-GPU heterogeneous computing, decomposing the domain in multiple dimensions can easily generate a number of small contacting faces which require frequent data transfers among neighboring blocks. One simple way to avoid this issue is to decompose the domain only in the $N_l$ dimension. If a number of $C$ CPUs are used to parallelize the residual calculation, the parallel execution time for the CPU version is given in Eq.~\ref{cpu}. Note that no matter how many parallel CPU cores are used, each decomposed block has a fixed amount of ghost cells in the decomposition dimension.

\begin{equation}
\label{cpu}
time_{gpu} = \frac{N_l N_w}{C} t_I + (2 N_w + 2 \frac{N_l}{C}) t_B + t_{DH} + t_{HH} + t_{HD} + 5 t_S
\end{equation}

For a scenario in which a number of $G$ GPUs are used, and suppose that the compute capability of a single GPU is $r_{gc}$ times stronger than a single CPU, the parallel execution time on multiple GPUs can be derived in Eq.~\ref{gpu}.

\begin{equation}
\label{gpu}
time_{gpu} = \frac{N_l N_w}{G r_{gc}} t_I + (2 N_w + 2 \frac{N_l}{G r_{gc}}) t_B + t_{DH} + t_{HH} + t_{HD} + 5 t_S
\end{equation}

Combining Eq.~\ref{cpu} and Eq.~\ref{gpu}, the execution time for a whole simulation applying CPU-GPU heterogeneous computing is given in Eq.~\ref{hete}.

\begin{equation}
\label{hete}
time_{hete} = \frac{N_l N_w}{G r_{gc} + C} t_I + (2 N_w + 2 \frac{N_l}{G r_{gc} + C}) t_B + t_{DH} + t_{HH} + t_{HD} + 5 t_S
\end{equation}

It can be seen that we may need to profile each test case to obtain the data transfer cost and the synchronization cost, which is difficult and unrealistic. In this work, the influence of the data transfer and synchronization are aggregated into a total variable, $t_f$, which should not be a constant. $t_f$ is affected by the problem size, number of processes, data transfer size, data transfer frequency, etc. Considering that the total data transfer cost is strongly affected by connected boundary conditions, we make an assumption that $t_f = \alpha T_b$, where $\alpha$ is a parameter and $T_b$ is the total time spent on the boundary enforcement. In this way, we can obtain Eq.~\ref{alpha}:

\begin{equation}
\label{alpha}
time_{hete} = \frac{N_l N_w}{G r_{gc} + C} t_I + (2 N_w + 2 \frac{N_l}{G r_{gc} + C}) t_B (1 + \alpha)
\end{equation}

In Eq.~\ref{alpha}, there are two different times $t_I$ and $t_B$, representing the execution time for interior domain and boundary domain, respectively. For a second-order finite volume CFD code, the relative ratio of $t_I$ over $t_B$ is fixed, so we have $t_B = \beta t_I$, where $\beta$ is in the range of 0.2$\sim$0.5. Using a profiling tool, the coefficient $\beta$ can be accurately obtained.

\begin{equation}
\label{beta}
time_{hete} = [\frac{N_l N_w}{G r_{gc} + C} + (2 N_w + 2 \frac{N_l}{G r_{gc} + C}) (1 + \alpha) \beta] t_I
\end{equation}

\section{Performance Metrics}
\label{metrics}

To compare the performance of a problem with different sizes on different computing platforms, the wall clock time per iteration step is converted to \emph{ssspnt}, which is scaled size steps per np time and defined in 
Eq.~\ref{ssspnt}.
\begin{equation}
\label{ssspnt}
\mathrm{ssspnt}=s\frac{size \times steps}{np \times time}
\end{equation}
where $s$ is a scaling factor to scale \emph{ssspnt} to the range of [0,1] on slow computing platforms, $size$ is the problem size denoting the number of finite cells, $steps$ is the total iteration steps, $np$ is the number of compute units and $time$ is the CFD solver wall clock time for $steps$ iterations. $s$ is chosen to be $10^{-6}$ in this work. 

\begin{figure}[H]
	\centering 
	\subfigure[Scaling behaviours using \emph{ssspnt}]{ 
		\label{ssspnt_scaling}
		\includegraphics[width=.45\textwidth,trim=5 5 5 5,clip]{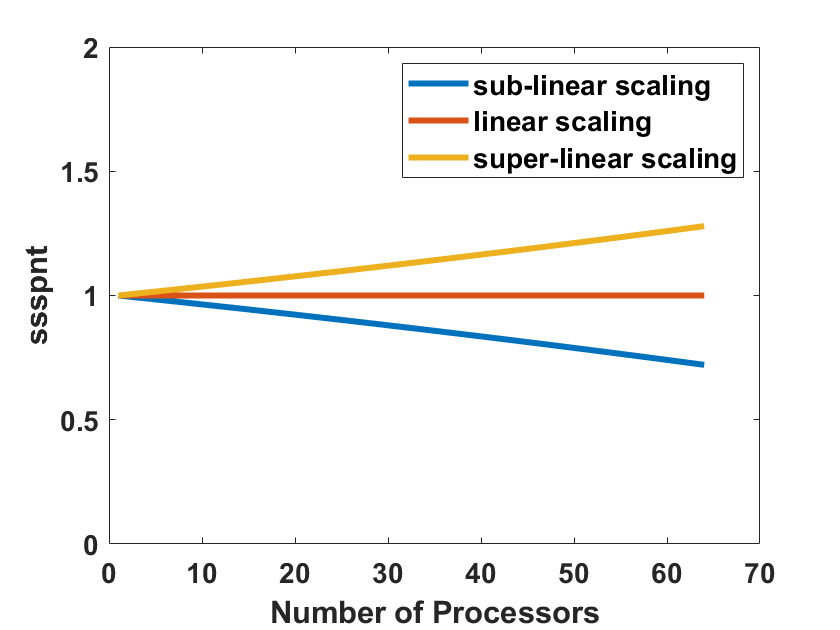} 
	}
	\subfigure[\emph{ssspnt} in different scenarios]{ 
		\label{ssspnt_scenarios}
		\includegraphics[width=.45\textwidth,trim=5 5 5 5,clip]{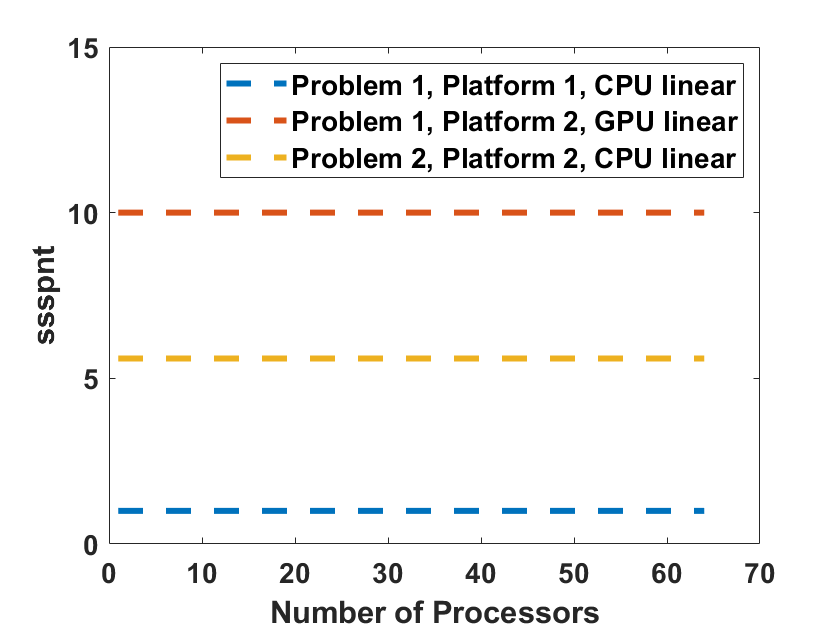} 
	} 
	\caption{Advantages of using \emph{ssspnt} as the performance metric} 
	\label{ssspnt_exp}
\end{figure}

It is easy to convert the metric \emph{ssspnt} to FLOPS if we know the operation count for each iteration. However, it would be difficult for us to manually count the total number of operations in complicated code, especially in real-world applications. \emph{ssspnt} is a better way for the performance evaluation compared to speedup or efficiency as \emph{ssspnt} also takes into account the problem size so that the performance for different problem sizes can be compared directly. \emph{ssspnt} enables us to better understand the relative speed difference than the metric "efficiency" in different scenarios with various problem sizes and platforms. Using \emph{ssspnt}, it is straightforward to know whether the scaling performance is super-linear, linear, or sub-linear, which is shown in Fig.~\ref{ssspnt_scaling}, as well as know the relative speed in different scenarios including the problems and platforms, which is shown in Fig.~\ref{ssspnt_scenarios}. 

Every node of the test platform in this work is equipped with two Intel Xeon E5-2680v4 (Broadwell) 2.4GHz CPUs, 512 GB memory, and two NVIDIA P100 GPUs. A maximum of 8 nodes are used. Each NVIDIA P100 GPU is capable of up to 4.7 TeraFLOPS of double-precision performance. The compilers used are PGI 18.1 and Open MPI 3.0.0. A compiler optimization of -O4 is used.

\section{Results and Discussion}
\label{results}

One CFD test case in this work is the supersonic flow through a simplified 2D 30-degree inlet. The CPU serial solution, multi-CPU parallel solution, GPU serial solution, multi-GPU parallel solution have been carefully compared and the relative difference among all the solutions is less than ($10^{-12}$).

The inflow conditions are given in Table~\ref{inlet_inflow}.

\begin{table}[H]
	\caption{Supersonic inflow boundary conditions}
	\centering
	\begin{tabular}{cc}
		\hline
		Mach number& 4.0\\
		Pressure& 12270 Pa\\
		Temperature& 217 K\\
		\hline
	\end{tabular}
	\label{inlet_inflow}
\end{table}

\begin{figure}[H]
	\centering 
	\subfigure[Mach number contour and streamlines]{ 
		\label{inlet_Ma}
		\includegraphics[width=.45\textwidth,trim=7 7 7 7,clip]{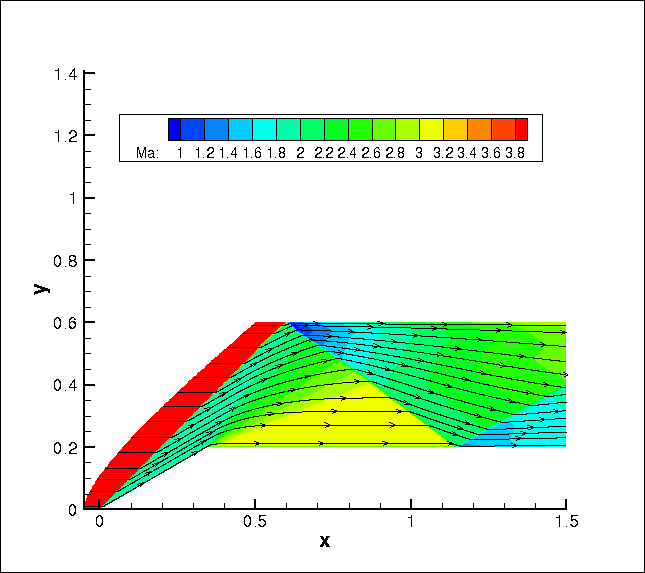} 
	} 
	\subfigure[Density contour]{ 
		\label{inlet_rho_euler}
		\includegraphics[width=.45\textwidth,trim=7 7 7 7,clip]{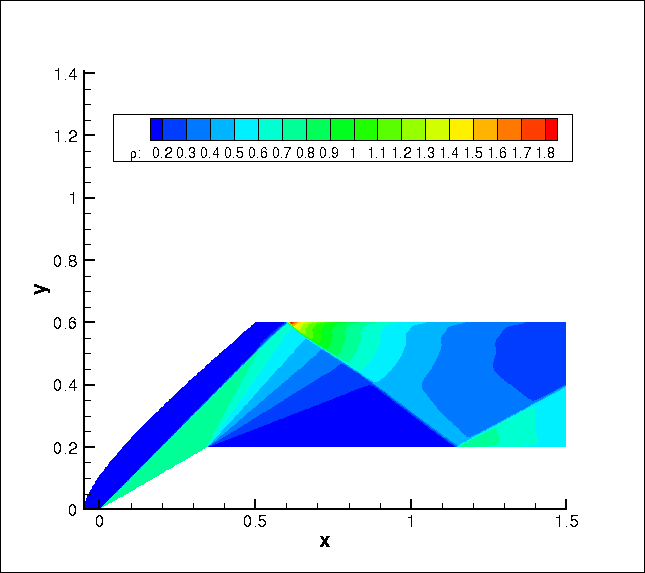} 
	} 
	\caption{Solution of 2D supersonic flow} 
	\label{Inlet-Mach-rho}
\end{figure}

The Mach number, streamlines and density contours are given in Fig.~\ref{Inlet-Mach-rho}.

There are some factors affecting the overall performance of CPU-GPU heterogeneous computing. First, the optimal single GPU worker over single CPU worker workload ratio needs to be determined, which can be found through running a series of cases having different workload ratios. As is known, the GPU worker compute much faster than the CPU worker, so the optimal workload ratio denotes the balance point between the CPU worker and the GPU worker. The optimal workload ratio is related to one single GPU over one single CPU speedup in their pure modes but is likely to be different, as the CPU worker needs to handle some computational work as well as some control work in the heterogeneous computing mode, adding synchronization overhead to the whole computation. Second, the number of GPU workers and the number of CPU workers may also affect the overall performance. Although a larger number of CPU workers seems to offer more computational power, the use of more processes may increase the communication overhead and synchronization overhead. We did numerical experiments for the Euler solver and NS solver using different single GPU over CPU workload ratios and different GPU and CPU combinations in this work.

\begin{figure}[H]
	\centering
	\includegraphics[width=.8\textwidth]{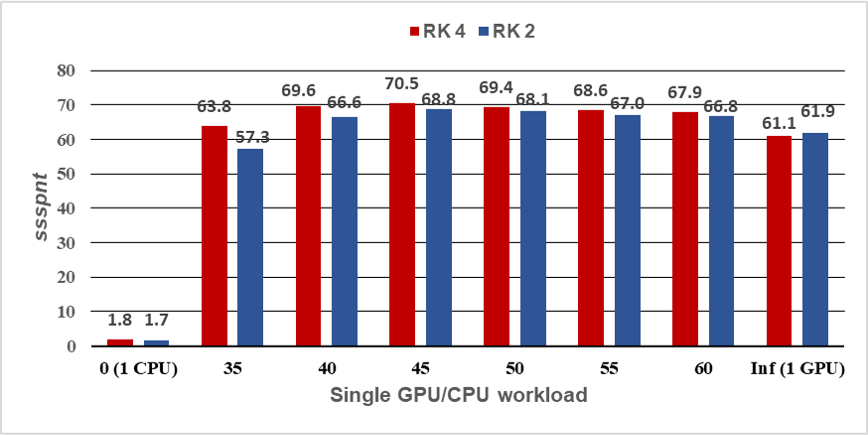}
	\caption{The \emph{SSSPNT} performance comparison between RK 2 and RK 4}
	\label{RK4_vs_RK2}
\end{figure}

Fig.~\ref{RK4_vs_RK2} shows the \emph{SSSPNT} performance comparison between RK 2 and RK 4 schemes in the SENSEI CFD code. It should be noted that the number of substeps is included in Eq.\ref{ssspnt}. Schemes having more intermediate steps tend to be slightly more efficient, possibly because more RK step calculations are more compact and inner sub-step iteration calculations have smaller synchronization overhead. Overall, the RK 2 and RK 4 temporal schemes have a similar operational intensity since the Flops per byte of DRAM traffic is similar in each substep. The performance difference is negligible so we can choose one RK scheme (RK 4 in this work) to represent a family of RK schemes.

\begin{figure}[H]
	\centering
	\includegraphics[width=.57\textwidth]{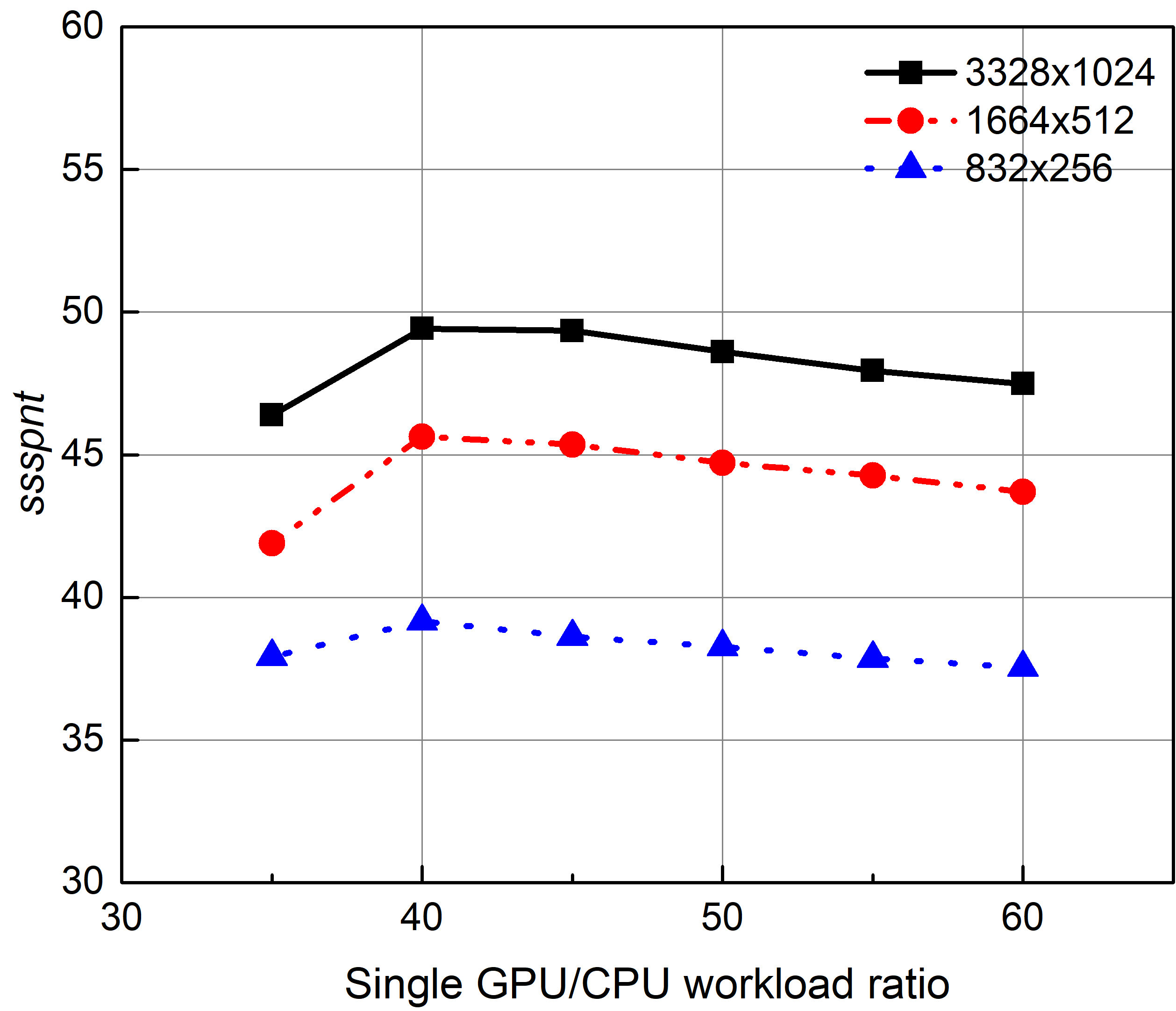}
	\caption{The influence of the problem size}
	\label{inlet_NS_large_vs_small}
\end{figure}

The influence of the problem size is presented in Fig.~\ref{inlet_NS_large_vs_small}. Generally, the \emph{SSSPNT} performance trends on various problem sizes are similar so that a global optimal single GPU over CPU workload ratio, around 40, can be obtained. It is easier to achieve higher \emph{SSSPNT} on larger problems as the GPU needs to be saturated to work efficiently. This problem can easily become memory bounded as small workloads are assigned to CPU workers, causing the number of processes to increase quickly and increasing the communication overhead. Simple domain decomposition such as 1D domain decomposition should be used to avoid complicated communication patterns on modern CPU-GPU heterogeneous systems in which the GPU usually runs much faster than the CPU.

\begin{figure}[H]
	\centering 
	\subfigure[The effect of GPU/CPU worker workload ratio]{ 
		\label{inlet_Euler}
		\includegraphics[width=.45\textwidth,trim=0 0 0 0,clip]{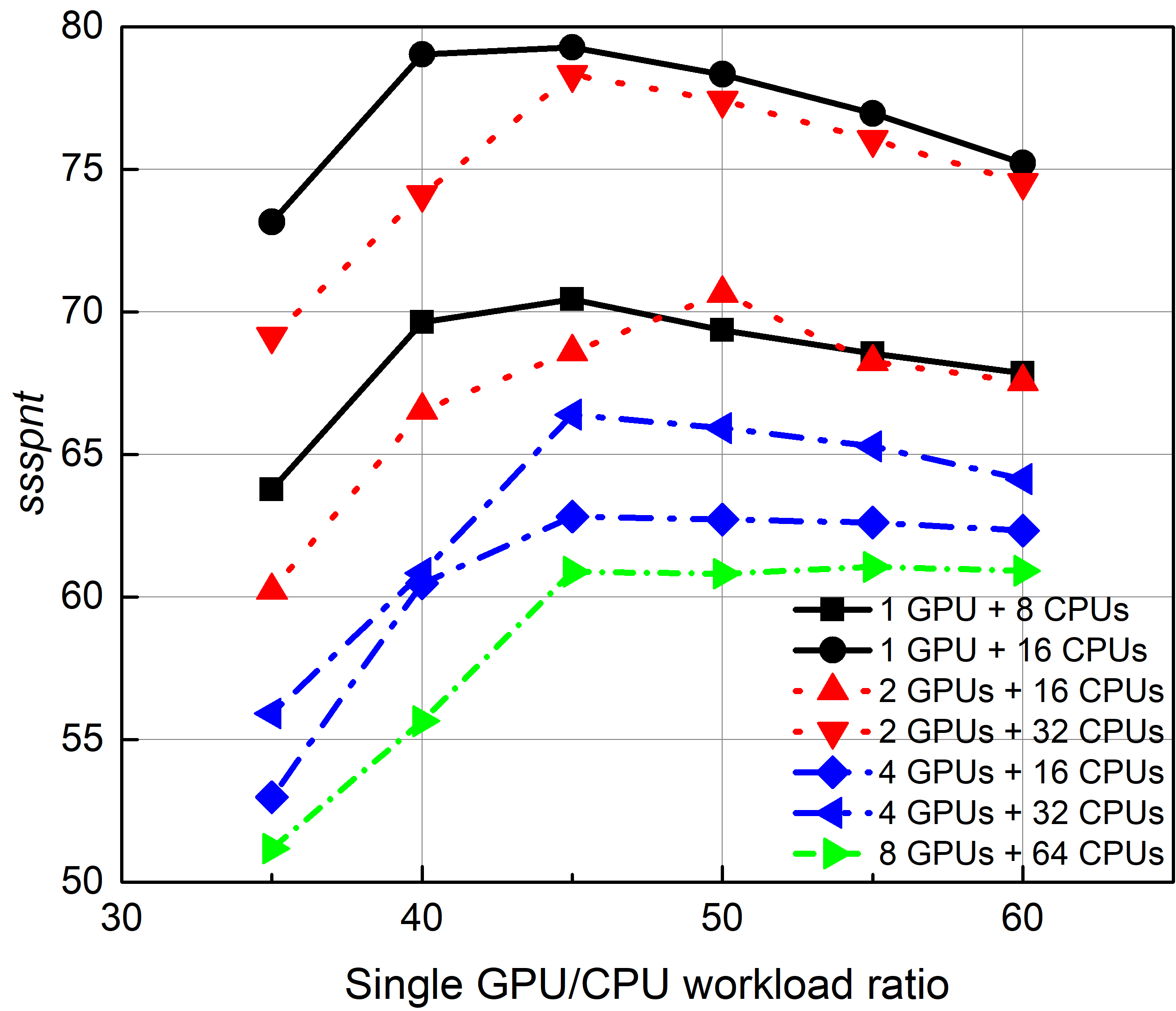} 
	} 
	\subfigure[Performance comparison of the heterogeneous computing and pure CPU/GPU computing mode]{ 
		\label{inlet_Euler_hete_vs_pure_vs_pred}
		\includegraphics[width=.45\textwidth,trim=0 0 0 0,clip]{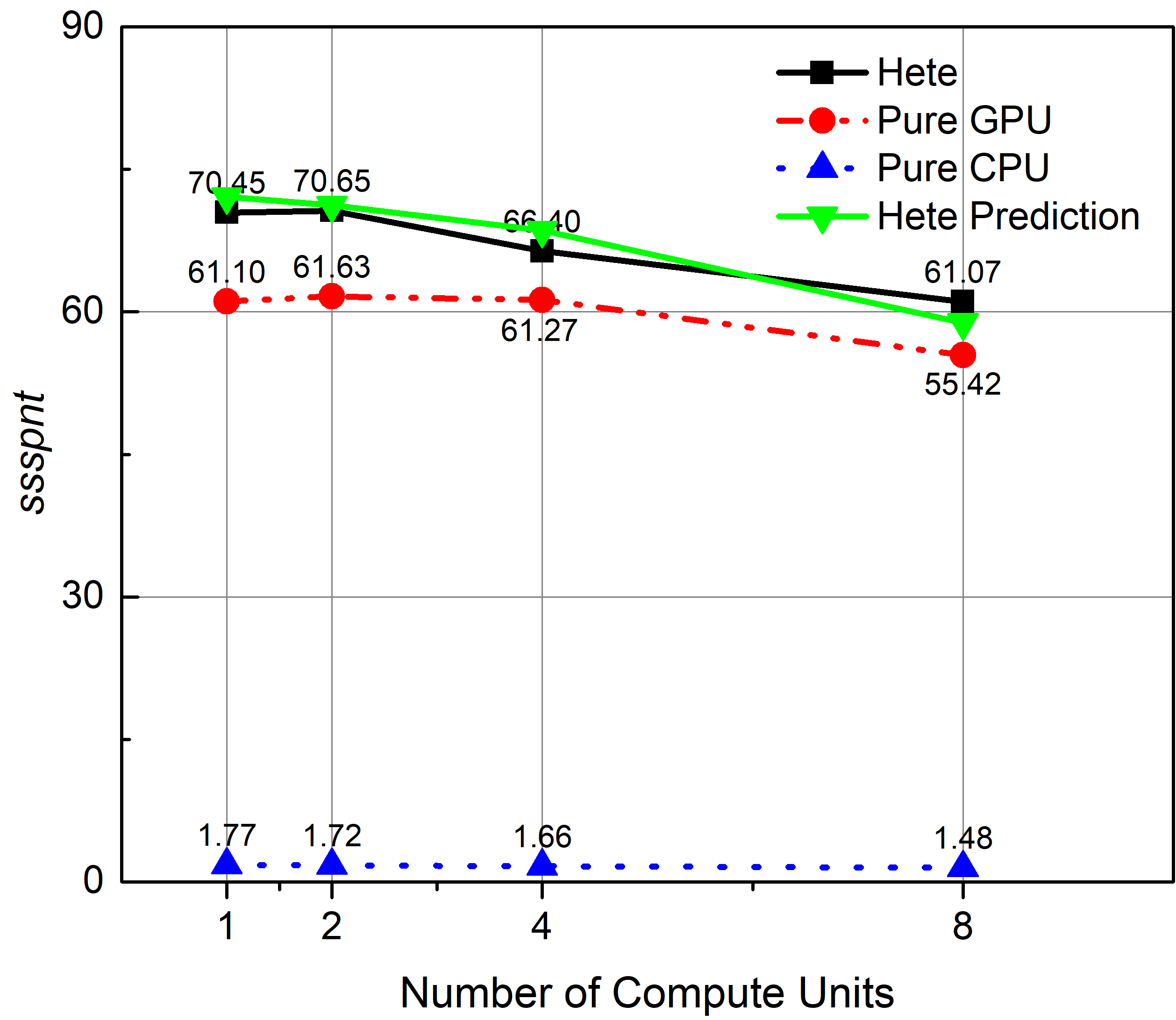} 
	} 
	\caption{\emph{SSSPNT} Performance of the Euler solver} 
	\label{hete_euler}
\end{figure}

Fig.~\ref{hete_euler} shows the performance of the Euler solver using \emph{ssspnt} in SENSEI. It should be noted that the variable \emph{np} in Eq.~\ref{ssspnt} is chosen to be the GPU number (not counting the CPU number) in the heterogeneous computing mode. Different numbers of GPU workers and CPU workers as well as different single GPU/CPU workload ratios are tested. In Fig.~\ref{inlet_Euler}, different curves denote the use of different numbers of CPU and GPU combinations. It can be found that the \emph{ssspnt} increases quickly when increasing the single GPU over single CPU workload ratio, which indicates that the workload ratio is quickly balanced between the CPU worker and the GPU worker. When increasing the workload ratio further, the performance becomes worse as the CPU worker contribution is negligible compared to the GPU worker contribution. An optimal single GPU over single CPU workload ratio can be obtained for each CPU-GPU combination. The optimal workload ratio range is between 40 and 55, which is close to the single GPU over a single CPU speedup in the pure mode. In Fig.~\ref{inlet_Euler}, there are some curves using the same number of GPUs but with different numbers of CPUs, for example, 1 GPU + 8 CPUs vs 1 GPU + 16 CPUs and 2 GPUs + 16 CPUs vs 2 GPUs + 32 CPUs. Adding more CPU workers increases the \emph{ssspnt}, especially when using a smaller number of GPUs. This can be explained that the communication cost increases when the number of processes increases, causing the benefits of heterogeneous computing to drop. Since the goal of applying heterogeneous computing is to run simulations faster than the pure computing mode, we can collect all the optimal data points and plot them together, which is the solid line with the square symbol in Fig.~\ref{inlet_Euler_hete_vs_pure_vs_pred}. One compute unit for the heterogeneous mode in Fig.~\ref{inlet_Euler_hete_vs_pure_vs_pred} denotes 1 GPU + 8 CPUs. As can be seen, the heterogeneous mode is faster than the pure GPU mode, although the performance gain is not significant. The heterogeneous computing mode may have worse scaling since the increased number of processes may incur greater communication overhead. The solid line with the lower triangle symbol represents the heterogeneous prediction based on the pure GPU and pure CPU data. The heterogeneous prediction is matched fairly well with the actual heterogeneous line, which proves the validity of the performance model for CPU-GPU heterogeneous computing in this work.

\begin{figure}[H]
	\centering 
	\subfigure[The effect of GPU/CPU worker workload ratio]{ 
		\label{inlet_NS}
		\includegraphics[width=.45\textwidth,trim=0 0 0 0,clip]{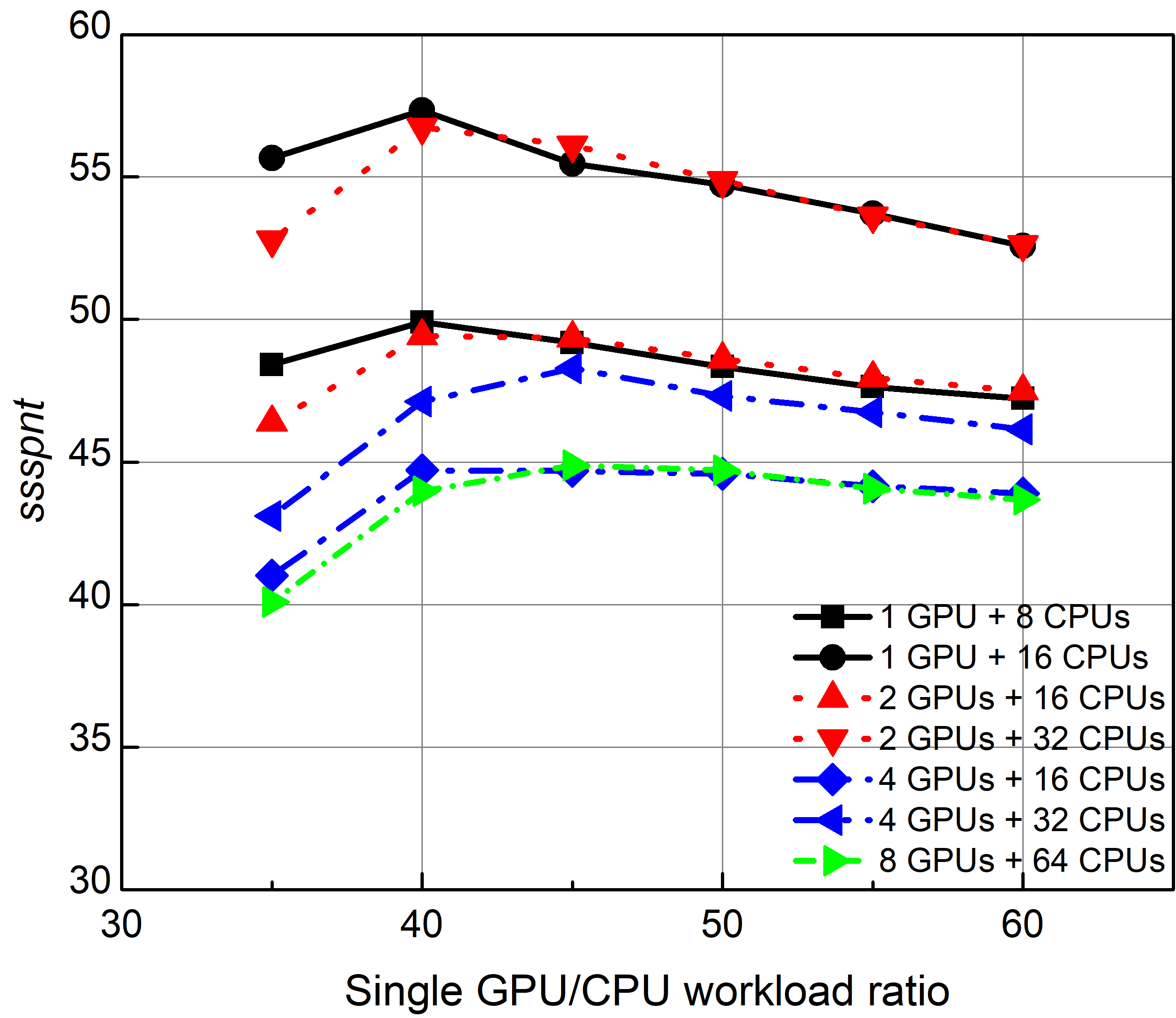} 
	} 
	\subfigure[Performance comparison of the heterogeneous computing and pure CPU/GPU computing mode]{ 
		\label{inlet_NS_hete_vs_pure_vs_pred}
		\includegraphics[width=.45\textwidth,trim=0 0 0 0,clip]{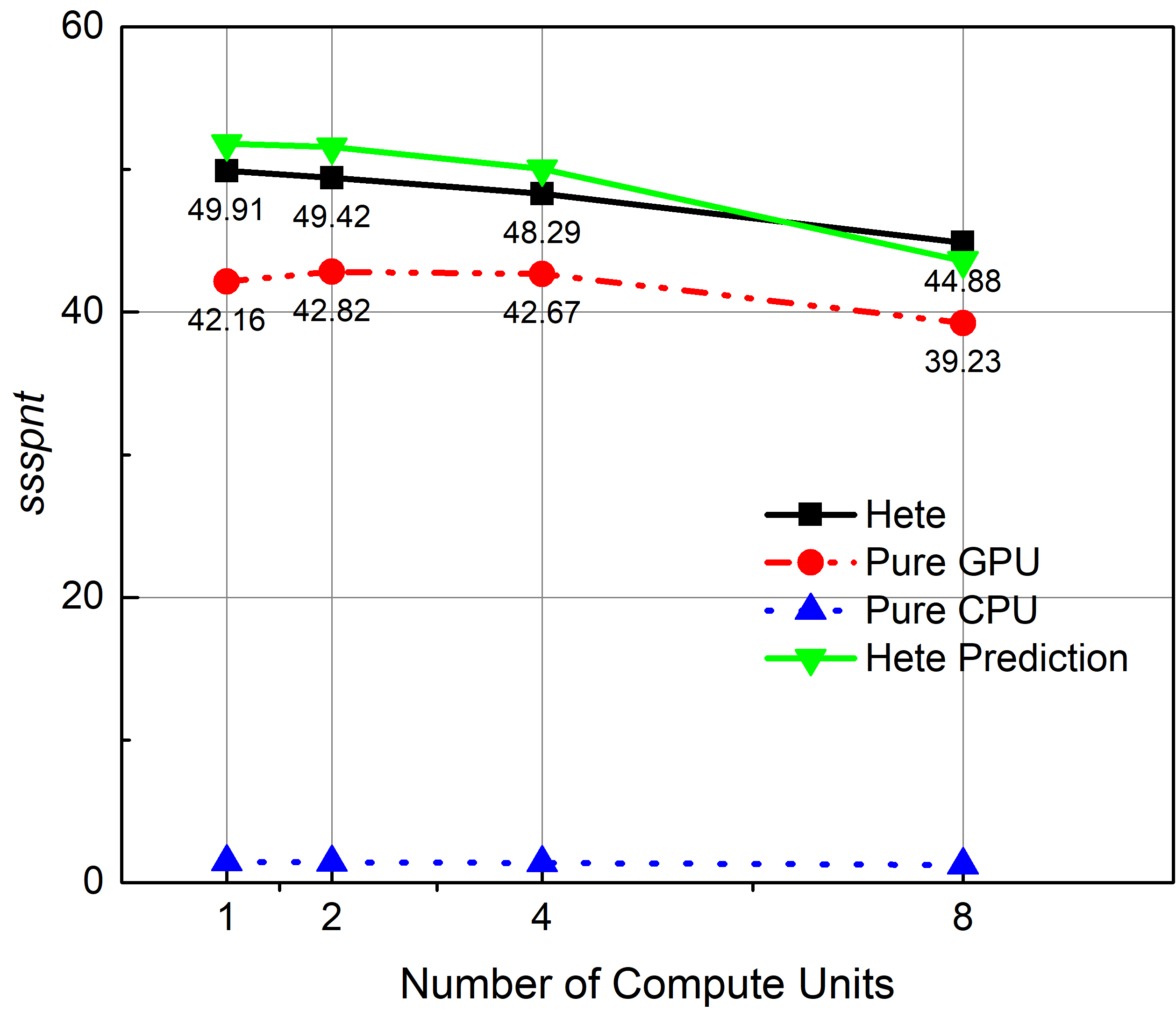} 
	} 
	\caption{\emph{SSSPNT} Performance of the NS solver} 
	\label{hete_ns}
\end{figure}

Fig.~\ref{hete_ns} shows the \emph{SSSPNT} performance of the Navier-Stokes solver in the SENSEI CFD code. The \emph{SSSPNT} values are smaller than those using the Euler solver as the Navier-Stokes solver involves the viscous flux calculation while the Euler solver does not. Overall, the trend of the workload ratio effect in Fig.~\ref{inlet_NS} and the performance comparison of different modes in Fig.~\ref{inlet_NS_hete_vs_pure_vs_pred} are generally similar to what can be summarized for the Euler solver in Fig.~\ref{hete_euler}. The performance model overpredicts the heterogeneous performance when the number of compute units is smaller, indicating that the effect of synchronization and data transfer cost is underestimated. However, the overall trend of the heterogeneous prediction is fairly reasonable. 

\section{Conclusions}
\label{conclusions}

In this work, a heterogeneous computing implementation for a finite volume CFD solver on a CPU-GPU system is presented. Its performance has been compared to a pure-GPU implementation and displays some performance advantages. The approach can be easily extended to other scientific computing fields to fully utilize heterogeneous computing resources. 

The heterogeneous computing implementation which utilizes a number of CPUs and GPUs can outperform the pure-GPU computing implementation by up to about 1.23 times. An efficient heterogeneous computing implementation requires the workload to be distributed reasonably among all co-execution units to match their own computing capacities, and it also requires the benefits of combining the total computational power of heterogeneous units to suppress the easily increased communication overhead. Future computer architects or application engineers may need to be aware of this when designing a new computer system or applying heterogeneous computing to accelerate their own applications.

The performance model for CPU-GPU heterogeneous computing proposed in this work can be used to estimate CFD applications requiring ghost data exchange. The performance model can also be extended to other scientific computing domains or help the design of future-generation heterogeneous systems.



%%       \citet{<label>} ==> Jones et al. [21]
%%       \citep{<label>} ==> [21]
%%

\bibliographystyle{elsarticle-num-names} 


%\begin{thebibliography}{00}



%\end{thebibliography}

\bibliography{mybib}

\end{document}